\documentclass[aps,prb,twocolumn,superscriptaddress,groupedaddress]{revtex4}

\usepackage{amssymb} \usepackage{color,graphicx} \usepackage{amsmath}
\usepackage{amsbsy} \usepackage{amsthm} \usepackage{bbm}
\usepackage{bm,bbm} \usepackage{float} \usepackage{braket}
\usepackage{placeins}
\usepackage[colorlinks=true,citecolor=blue,linkcolor=red,urlcolor=red]{hyperref}
\usepackage[sort&compress]{natbib}
\usepackage{comment}
\usepackage{color}
\usepackage{braket}

\newcommand{\bq}{\begin{equation}} \newcommand{\eq}{\end{equation}}
\newcommand{\bqali}{\bq\begin{aligned}}
\newcommand{\eqali}{\end{aligned}\eq}
\newcommand{\bqn}{\begin{equation*}}
\newcommand{\eqn}{\end{equation*}}

\newcommand\D{\operatorname{d}}
\renewcommand\k{{\bf k}}

\renewcommand\r{{\bf r}}

\graphicspath{{../Figures/}}

\begin{document}

\title{When Cavendish meets Feynman: A quantum torsion balance for testing the quantumness of gravity
}

\author{Matteo Carlesso}
\email{matteo.carlesso@ts.infn.it}
\affiliation{Department of Physics, University of Trieste, Strada Costiera 11, 34151 Trieste, Italy}
\affiliation{Istituto Nazionale di Fisica Nucleare, Trieste Section, Via Valerio 2, 34127 Trieste, Italy}
\author{Mauro Paternostro}
\affiliation{Centre for Theoretical Atomic, Molecular and Optical Physics, School of Mathematics and Physics,
Queen's University Belfast, Belfast BT7 1NN, United Kingdom}
\affiliation{Laboratoire Kastler Brossel, ENS-PSL Research University, 24 rue Lhomond, F-75005 Paris, France}
\author{Hendrik Ulbricht}
\affiliation{Department of Physics and Astronomy, University of Southampton, SO17 1BJ, UK}
\author{Angelo Bassi}
\affiliation{Department of Physics, University of Trieste, Strada Costiera 11, 34151 Trieste, Italy}
\affiliation{Istituto Nazionale di Fisica Nucleare, Trieste Section, Via Valerio 2, 34127 Trieste, Italy}

\date{\today}
\begin{abstract}
We propose a thought experiment, based on a mechanism that is reminiscent of Cavendish's torsion balance, to investigate the possible quantum nature of the gravitational field generated by the quantum superposition state of a massive system. Our proposal makes use of the dynamics of a ultra-stable optically levitated nanomechanical rotor endowed with a spin to generate a quantum angular superposition that is then tested through standard Ramsey-like scheme. Gravity manifests itself as an effective decoherence mechanism, whose strength is different--and, as we show, appreciable--in the classical and quantum case. By incorporating both the source for decoherence and the mechanism to probe it, the experiment that we propose allows for a much reduced degree of control and dynamical engineering.
\end{abstract}
\pacs{} \maketitle

In his {\it Lectures on Gravitation}~\cite{FeynmanGravitation} Feynman famously wondered: {\it ``Is it possible that gravity is not quantized and all the rest of the world is?"}. He then proceeded to propose a thought experiment that, if realised, would show the quantum nature of gravity.

However, he also considered the possibility that quantum theory fails beyond a given scale of mass, distance or complexity, ascribing to gravity the possible cause of such a failure, a viewpoint later reprised by Penrose, who put forward the concept of gravity-induced collapse of the wavefunction~\cite{PenroseCollapse}.

As of today there is no experimental evidence of the possible quantum nature of gravity, nor a unanimous consensus on the potential features of a theory of quantum gravitation. We are not even able to answer the simple question: would the gravitational field resulting from a system prepared in a quantum superposition be itself a superposition of two fields, as one would expect in analogy to electrodynamics? It clearly appears that the questions posed by Feynman are more pressing than ever, as witnessed by the substantive body of literature addressing them~\cite{hoss,arun,kiefer,bassi,bahrami,kafri}. 

Here we propose an experimentally viable scheme for engineering the spatial superposition of a massive system, generating an appreciable gravitational field. Our proposal, which is based on an optomechanical platform, creates a coherent superposition of distinguishable states of a {\it torsional degree of freedom} of a nanorod.
Two possibilities are checked. In the first case the gravitational field is itself in a superposition, preserving by linearity the quantum state of the system. In the second case the gravitational field is classical, equally distributed between the two states of the system, generating a mutual attraction, thus reducing the angular distance.

This is reminiscent of the gravity experiment by Cavendish~\cite{Cavendish}, where a torsion balance was used to investigate gravitational effects.  Analogously, the exquisite sensitivity of our proposed {\it quantum torsion balance} allows us to discriminate the nature of the gravitational field generated by a quantum superposition, thus providing evidences in favour or against the quantumness of gravity. Our scheme thus represents the optomechanical embodiment of Feynman's thought experiment, built upon Cavendish's intuition. As the system that we propose would be capable of self-testing the character of the the gravitational field generated by the quantum superposition, without the need for an external probe, our scheme appears to offer significant advantages over previously proposed ideas for testing the nature of gravity.

\begin{figure}[b!]
\includegraphics[width=0.9\linewidth]{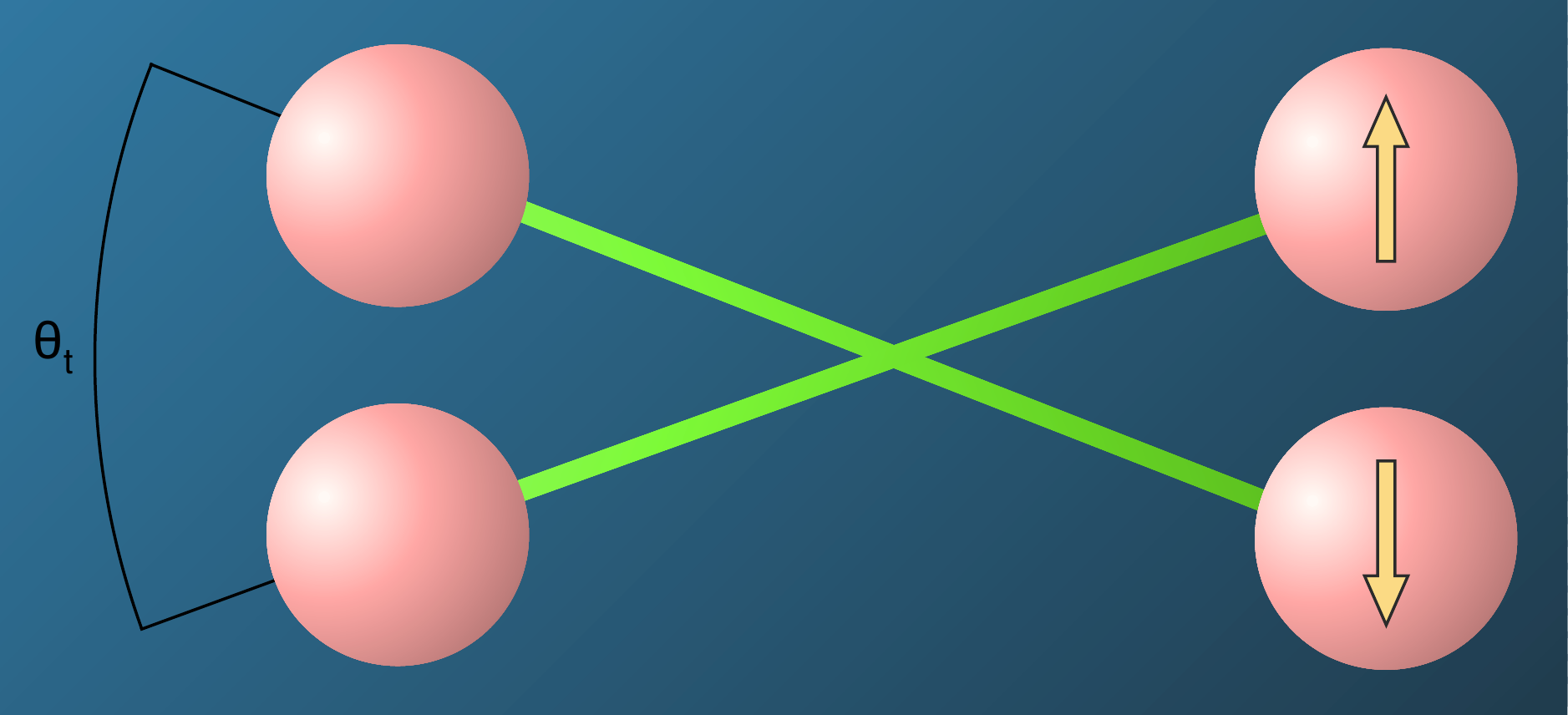}
\caption{Graphical representation of the nanorod (depicted with two pink spheres and a green bar) in angular superposition. The angular superposition is consequence of the spin superposition (yellow arrows), and it is initially prepared at an angular distance $\theta_0$. According to the classical gravity scenario they attract each other, and the angular distance $\theta_t$ reduces in time.}
\label{fig-1}
\end{figure}

\noindent
{\bf The system.--} We consider an angular superposition of a non-spherical object, as represented in Fig.~\ref{fig-1}, which is modeled by two spheres of mass $m$ and radius $r$ connected to each other by a rigid bar of negligible mass and length $2L$. We refer to such a system as a {\it nanorod} from now on. Any other non-spherical system with similar dimensions is expected to give a contribution of the same order of magnitude. We assume the gravitational attraction as given by Newton's law, which so far proved to be fully adequate to describe the behaviour of matter in the non-relativistic regime. 

To generate the angular superposition we consider one of the two spheres to be endowed with a magnetic spin. The generation scheme thus follows the lines of a pulsed Ramsey technique~\cite{Scala:2013aa}. The spin can be provided, for instance, by a single spin-1 nitrogen-vacency (NV) centre in diamond.\\
\noindent
{\bf The proposed experiment.--} We now describe the proposed experimental procedure to prepare the rotational superposition state and measure the effect of gravity. We also estimate the values of the key parameters of the setup, which are all found to be reachable by state-of-the-art experimental capabilities. 

\noindent
{\it 1) Cooling the torsional motion of the nanorod--} Levitated optomechanics provides a variety of experimental techniques to optically trap and prepare rotational states of optically trapped non-spherical particles \cite{Arita2013, Kuhn2015, Kuhn2017, Hoang2016, Rahman2017}. Our proposed experiment requires the cooling of the torsional motion of the nanorod to temperatures close (but not necessarily equal) to those equivalent to its rotational quantum ground state. Such working point is achievable by current experiments based on the translational motion of levitated optomechanics~\cite{jain2016direct} and, through  continuous feedback, the free rotation of a nanorod~\cite{Kuhn2017b}. We stress that achieving exactly the ground state energy of the torsional mode at hand here is not necessary for our purposes, thus reducing significantly the difficulties of the scheme. \\
\noindent  
{\it 2) Generation of a spin superposition--} After cooling the torsional motion of the nanorod to a low phonon state, a microwave (MW) $\pi /2$-pulse is used to generate a superposition state of the spin of the form $(\ket{+1}+\ket{-1})/\sqrt{2}$, where $\ket{\pm1}$ are the two spin states. This is a standard and mature technique in magnetic resonance experiments \cite{Levitt2001}. The important parameter for our scheme is the lifetime of the generated spin state. The decoherence time $T_2$, for magnetic resonance spin experiments is usually estimated $\sim$300 $\mu$s for diamond nanorods of diameter 300--500 nm, as experimentally confirmed by spin-echo techniques~\cite{Andrich}. For nanorods of 50 nm of diameter and 150 nm of length, a value of $T_2 = 80 \mu$s has been measured \cite{Trusheim2014}. We thus assume $T_2 = 100 \mu$s, which sets the upper limit to the next stage of our scheme.\\
\noindent
{\it 3) Transfer of the spin superposition state to the torsional motion of the nanorod: --} We switch on a non-homogeneous magnetic field whose gradient couples the motion of the nanorod to the spin. For a gradient of $\partial_x B=10^6$ T/m, which is one order of magnitude smaller than the maximum achieved by fine needle tips of permanent magnetic material~\cite{Schirhagl,Mamin}, a nanorod of length $L=10\,\mu$m and mass $m\sim10^{-20}$ kg, we find a separation angle of the superposition states of $\theta_0=7.92\times10^{-4}$\,rad when the magnetic field is kept on for 2.5\,$\mu$s, which is a time 40 times smaller than $T_2$. Note that this initial configuration is such that each sphere and its superimposed self do not overlap, both in terms of their size and of the size of their center-of-mass wave function. This eliminates a possible ambiguity in the classical case, on how gravity might act when there is an overlap in the superposition.\\
\noindent
{\it 4) Decoupling of spin-angular superposition--} The transfer of the spin superposition state to the torsional degree of freedom of the nanorod is completed by a procedure that decouples the spin state from the torsional mode. This is accomplished by measuring the state of the spin onto a superposed basis. This guarantees that the established coherence in the state of the joint spin-torsional system is {\it passed} entirely to the external degree of freedom of the nanorod and the latter becomes insensitive to any decoherence mechanisms of the spin.  \\
\noindent
{\it 5) Free evolution--} After the superposition-transfer stage, both the harmonic potential and the magnetic field are switched off, to let the nanorod evolve freely. 
 Only gravity, if classical, would force the rotational superposition state to change, as there is no competing quantum interaction.  We estimate that after a free evolution time of 2.5\,s, the superposition angle change is of the order of $|\theta_t-\theta_0|\simeq 5\times10^{-10}$\,rad if gravity is classical (and does not change if it is quantum). Such a long free evolution time can be achieved in a drop tower. Competing effects affecting the stability of the quantum superposition are estimated later on in this work and shown to be controllable. 
\\
\noindent
{\it 6) Detection of angular state of the nanorod--} The free evolution of the superposition is only interrupted by the measurement of the angular configuration of the nanorod. This is performed by switching back on the optical trap, which is falling together with the particle. The angle resolution has been shown experimentally to be on the order of less than $10^{-15}\,$m$/L=10^{-10}$\,m \cite{Vovrosh2017,Aspelmeyer2014}. This value can be used to estimate the here described protocol and is sufficient to resolve the desired effect of gravity. \\
\noindent
{\bf Theoretical analysis.--} We now provide the theoretical support for the proposed experimental scheme. 
For the transfer of the superposition we can take advantage of the following scheme\cite{Wan2016}. 
A uniform magnetic gradient $\partial_x B$ is introduced, and the untrapped sphere containing the NV center evolves according to the following Hamiltonian
\bq
\hat H=\frac{\hat p^2}{2m}- g_\text{\tiny NV}\mu_\text{\tiny B} (\partial_x B)\hat S_z\hat x,
\eq
where $m$ is the mass of the system, $g_\text{\tiny NV}\sim 2$ is the Land\'e factor of the NV center and $\mu_\text{\tiny B}$ is the Bohr magneton, $\hat S_z$ is the spin operator of the NV center spin along the $z$ direction,  $\hat x$ and $\hat p$ are the position and momentum operators along the $x$ direction. We assume the center of mass of the sphere to be in a coherent state. According to the above Hamiltonian, after a time $t_0$, the superposition of the spin is transferred to the state of the angular configuration\cite{Wan2016}. Thus, the total state of the nanorod reads:
 \bq
 \ket{\Psi(t_0)}=\frac{\ket{\psi(t_0,+1)}\ket{+1}+\ket{\psi(t_0,-1)}\ket{-1}}{\sqrt{2}}.
 \eq
Here, $\ket{\psi(t_0,\pm1)}$ represent the states of the angular degrees of freedom, whose angular separation is given by $\theta_0=\arcsin\left(\frac{\Delta_0}{2L}\right)$, where 
\bq
\Delta_0=\frac{g_\text{\tiny NV}\mu_\text{\tiny B}}{m}(\partial_x B)t_0^2.
\eq
The decoupling of spin from torsional state is now achieved by measuring the state of the former over the basis of the 
transversal magnetisation component $\hat S_x$, and post-selecting only the cases associated with the projection onto states $(\ket{+1}\pm\ket{-1})/\sqrt2$. This delivers the torsional state 
\begin{equation}
\ket{\Psi_{\rm tor}}=\frac{\ket{\psi(t_0+\delta t,+1)}\pm\ket{\psi(t_0+\delta t),-1}}{\sqrt2},
\end{equation}
where $\delta t\ll t_0$ is the time taken for the measurement to take place, and the relative phase between the distinguishable torsional states is inessential for our scopes.  
\begin{figure*}[th!]
\centering
\includegraphics[width=0.9\linewidth]{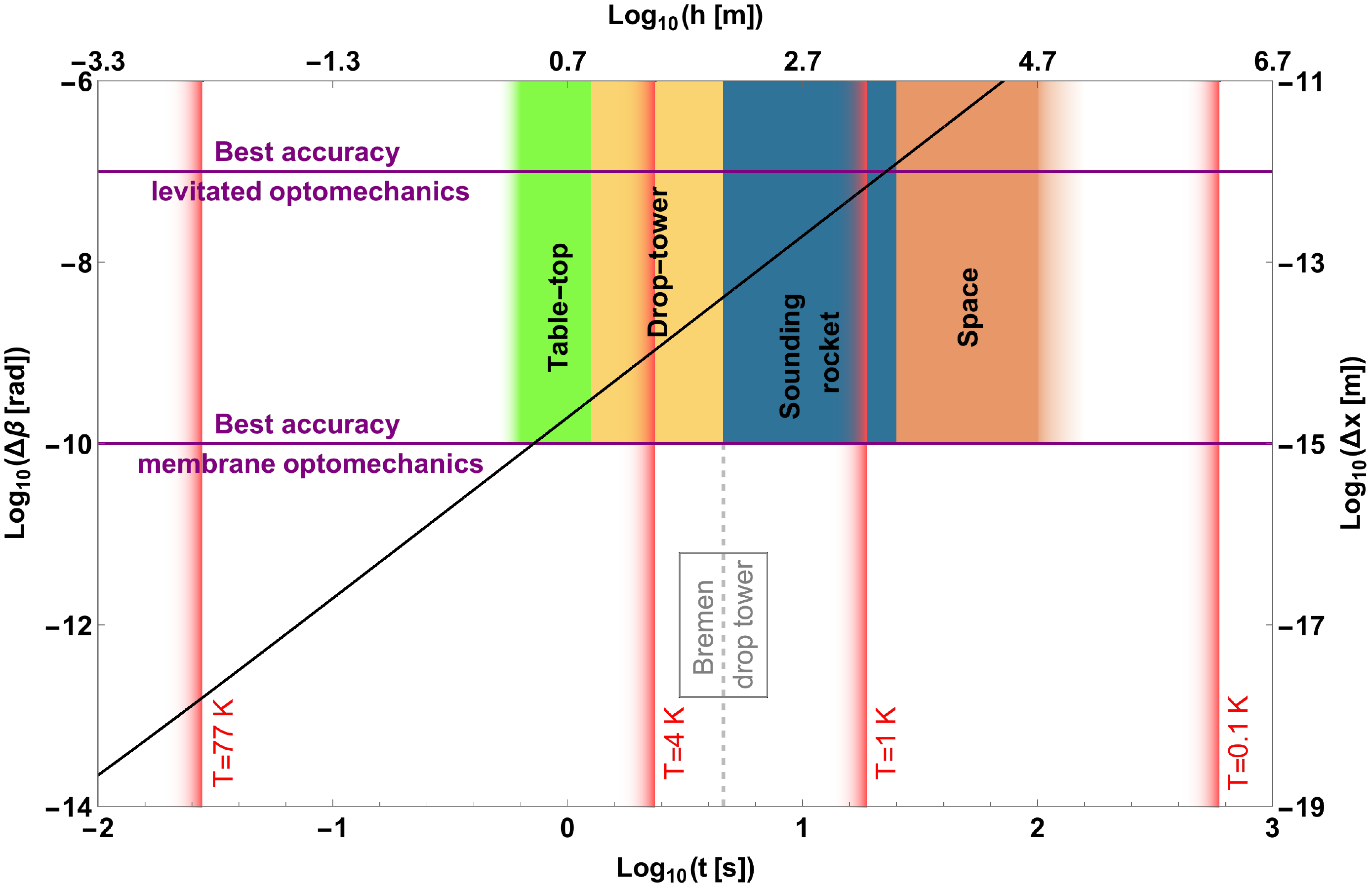}
caption{(\textit{Color online}) The black line shows the time evolution of the angular distance $|\theta_t-\theta_0|$ in the classical scenario. (In the quantum scenario, it does not change in time.) The chosen system is made of diamond (dielectric constant $\epsilon=5.7+2.85\times 10^{-4}i$) whose dimensions are $r=7.92$\,nm and $L=10\,\mu$m. By applying a magnetic gradient field of $10^6\,$T/m for $10\,\mu$s we obtain an initial superposition angle $\theta_0=7.92\times10^{-4}$\,rad. The distance $|\theta_t-\theta_0|$ is compared with the precision necessary to detect the motion: this is given in terms of radiants ($\Delta\theta$) and of meters ($\Delta x$), respectively, on the left and right vertical axes. The two purple horizontal lines denote the accuracy achieved in levitated\cite{Vovrosh2017} and in membrane optomechanics\cite{Aspelmeyer2014}. The decoherence times for different characteristic temperatures are reported in red: 77, 4, 1 and 0.1\,K. They were computed with reference to the maximal initial distance $\theta_0$, thus slightly overestimating the effect. The top horizontal axis indicates the distance $h$ of the free fall in gravity. The drop tower in Bremen ($4.6$\,s) is identified by the grey dashed line.}
\label{figres}
\end{figure*}

Next we quantify the difference induced according to the classical description of the gravitational field as compared to the quantum description, during the free evolution. In the quantum scenario, due to linearity, each part of the superposition does not feel the other part's gravitational potential. Therefore, the quantum superposition is ideally stable and the initial angular distance $\theta_0$ does not change in time. In the classical scenario, on the other hand, one needs to account also for the gravitational field generated by one part of the superposition and acting on the other. This results in a mutual attracting effect, and a corresponding reduction of the angular distance $|\theta_t-\theta_0|$. Such a reduction is the quantity to address in order to distinguish between the two descriptions of gravity. The full theoretical analysis is reported in the Supplementary Information. Fig.~\ref{figres} displays the distance $|\theta_t-\theta_0|$ as a function of time (black line). For comparison we report the best accuracies achieved in levitated optomechanics and membrane optomechanics (purple lines). The experiment can be performed as a table-top, drop-tower, sounding rocket or space experiment, and the respective timescales are  highlighted in green, grey, blue and orange respectively.

To detect the difference between the quantum and classical case, one needs to estimate decoherence effects, which potentially deteriorate the superposition. 
While a detailed calculation is deferred to SI, here we simply report the most preponderant mechanisms, and the associated decoherence rates~\cite{Schlosshauer:2007aa,{Romero-Isart:2011aa}}. First, we should consider collisions of the nanorod with the residual gas surrounding the system. The  decoherence rate for collisional decoherence reads [cf. SI]
\bq\label{coll}
\Lambda_\text{\tiny D}^\text{\tiny coll}=\frac{64n_\text{\tiny gas}\sqrt{2\pi m_\text{\tiny gas}}}{3\hbar^2}r^2L^2(k_\text{\tiny B}T_\text{\tiny E})^{3/2}\sin^2\tfrac\theta2,
\eq
where $m_\text{\tiny gas}$ is the mass of the residual gas, $T_\text{\tiny E}$ is its temperature, $n_\text{gas}$ is the gas density, $\zeta(x)$ is the Euler-Riemann zeta function of argument $x$, $k_\text{\tiny B}$ is the Boltzmann constant.

The second relevant mechanism of  decoherence is due to the scattering, absorption and emission of environmental thermal photons. We call $\Lambda^\text{\tiny i}_\text{\tiny D}~(\text{i=scatt,\,abs,\,em})$ the corresponding decoherence rates, which are [cf. SI]
\bq\label{scatt-9}
\begin{aligned}
&\Lambda_\text{\tiny D}^\text{\tiny scatt}=\frac{64\cdot 8!\xi(9)}{9\pi}r^6L^2c\, \Re\left(\frac{\epsilon-1}{\epsilon+2}\right)^2\left(\frac{k_\text{\tiny B}T_\text{\tiny E}}{\hbar c}\right)^9\sin^2\tfrac\theta2,\\
&\Lambda_\text{\tiny D}^\text{\tiny em (ab)}=\frac{128\pi^5}{189}\Im\left(\frac{\epsilon-1}{\epsilon+2}	\right)c\, r^3 L^2 \left(\frac{k_\text{\tiny B}T_\text{\tiny I(E)}}{\hbar c}\right)^6\sin^2\tfrac\theta2,
\end{aligned}
\eq
where $T_\text{\tiny I}$ is the internal temperature of the system, which in principle differs from that of the gas, for systems that are not in thermal equilibrium, and $\epsilon$ is the complex dieletric constant of the two spheres exemplifying the nanorod. 
The overall characteristic time for decoherence effects is $\tau_\text{\tiny D}=1/\Lambda_\text{\tiny D}$,  with $\Lambda_\text{\tiny D}=\tau^{-1}_\text{\tiny D}$ and $\Lambda_\text{\tiny D}=\sum_i \Lambda_\text{\tiny D}^{i}$.

In order to fix the ideas, we take the residual gas to be a mixture of nitrogen N$_2$ at $78\%$ and oxygen O$_2$ at $22\%$ at the  density of $10^9$\,particles/m$^3$, which corresponds to a pressure of $4\times10^{-14}$\,mbar at room temperature. Moreover we consider the equilibrium situation where $T_\text{\tiny E}=T_\text{\tiny I}$.

We report in Fig.~\ref{figres} the decoherence times for four typical temperatures in experiments: 77, 4, 1 and 0.1 \,K (red vertical lines). To avoid environmental disturbances on the free evolution, one must perform the experiment before decoherence dominates. Thus, an experiment performed at 300\,K should be concluded within a time of 0.0036\,s, while for 1\,K one can let the system evolve up to $\sim$19\,s.

\noindent
{\bf Discussion.--} We have proposed a quantum version of Cavendish torsional balance to assess the potential difference in the nature of the gravitational field produced by a coherent quantum superposition. Our scheme incorporates in the system to address both the source of the effect to discriminate, and the probing mechanism, relaxing the requirements for the implementation of the proposal itself. The use of the angular degree of freedom, rather than the translational one that is usually called for in levitated optomechanical setups, offers advantages of enhanced sensitivity and precision that put our proposal within reach of state-of-the-art technology. We believe that such a change in perspective could be of substantial help towards the implementation of an experiment to test the quantum nature of gravity.\\

\newpage

\section*{Supplementary Information}

\subsection{Calculation of the time evolution of the angular displacement in the classical scenario}

\begin{figure}[b!]
\includegraphics[width=1\linewidth]{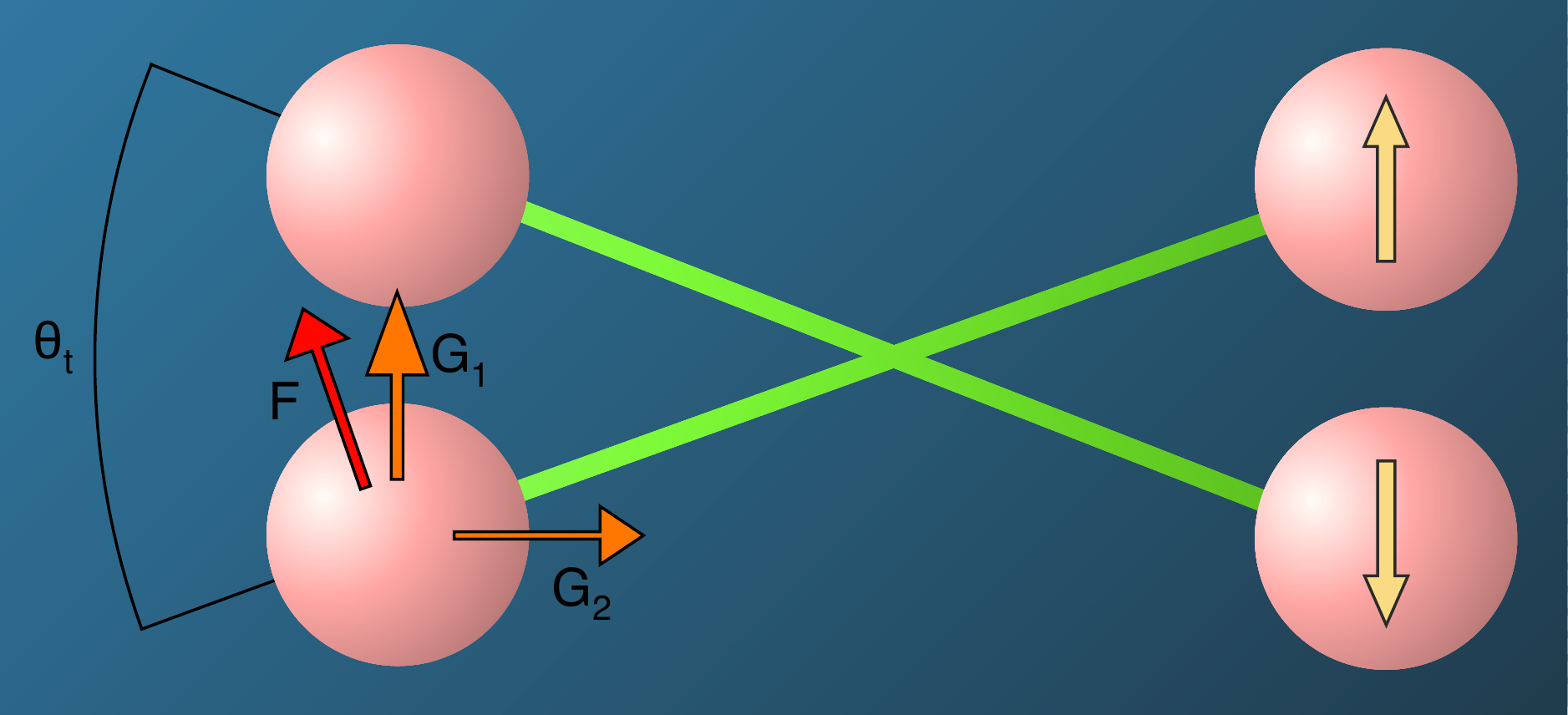}
\caption{(\textit{Color online}) Graphical representation of the nanorod in the classical gravity scenario. The Newtonian gravitational forces $G_1$ and $G_2$ acting on the bottom-left sphere are reported in orange. The tangent force $F$ which determines the angular acceleration is shown in red.}
\label{fig-1SI}
\end{figure}

Figure~\ref{fig-1SI} shows the gravitational forces attracting the two terms of the angular superposition.  Because of the rigid bar, the net force along the radial direction  is zero. Then, we need considering only on the tangent component.
The sum of the two forces ${\bm G_1}$ and ${\bm G_2}$ acting, for example, on the bottom-left mass along the tangent direction, is given by 
\bq
F=G_1 \cos\tfrac\theta2-G_2\sin\tfrac\theta2,
\eq
pointing clock-wise, and the angular acceleration is given by $\ddot\theta(t)=F/(mL)$. The magnitude of the two forces is given by the Newtonian expression $G_i= G m^2 / d_i^2$.
The distances with respect to the other masses are $d_1=2L \sin\tfrac\theta2$ and $d_2=2L \cos\tfrac\theta2$. 
The angular acceleration, in suitably rescaled units of time $t=\tau s$, is:
\bq\label{thetaddt}
\ddot \theta(s)=\frac{\cos^3\tfrac{\theta(s)}2-\sin^3\tfrac{\theta(s)}2}{\cos^2\tfrac{\theta(s)}2\sin^2\tfrac{\theta(s)}2}.
\eq
where $\tau$ is the characteristic time of the motion:
\bq\label{eq:tau}
\tau=\sqrt{\frac12\frac{4L^3}{Gm}}.
\eq
The factor $1/2$ in the definition of $\tau$ is due to the geometry of the system, as there are two equal masses contributing to the acceleration. The acceleration diverges when $\theta$ approaches 0 or $\pi$. These  cases correspond to when the masses are on top of each other, and the Newtonian potential becomes singular:  the equations obviously fail to apply. This situation is excluded in the experimental scheme here proposed. The acceleration is equal to zero when $\theta=\tfrac\pi2$, when the tangent forces balance each other.

\subsection{Analysis of decoherence rates}

To resolve the time scale parameter $\tau$, one has to compare it with the decoherence time $\tau_{\text{\tiny D}}$ which characterizes the life of the superposition.
There are several mechanisms of decoherence, the most relevant ones being collisions with the residual gas in the vacuum chamber; and scattering, absorption and emission of thermal photons. This is what we discuss now. 

The decoherence rate $\Lambda_\text{\tiny D}=\tau^{-1}_\text{\tiny D}$ is the sum $\Lambda_\text{\tiny D}=\sum_i \Lambda_\text{\tiny D}^{(i)}$ over the  decoherence mechanisms and $ \Lambda_\text{\tiny D}^{(i)}$ are given by \cite{Zhong:2016aa,Schlosshauer:2007aa} 
\bqali\label{lambdaD}
 \Lambda_\text{\tiny D}^{(i)}=&\int \D^3\k\int \frac{\D^2\hat {n}'}{4\pi} v(k) \mu(k) |f(\k,k\hat n')|^2\left(1-R(\k,\k')	\right),
\eqali
where $\k'=k\hat n'$, $\mu(k)$ is the momentum-space density of the  environmental particles, $v(k)$  their velocity.  $R(\k,\k') =   f(\k'_\theta,\k_\theta) / f(\k',\k)$ is  the ratio between two  scattering amplitudes: that referring to the rotated system ($\k_\theta$ corresponding to $\k$ rotated by an angle $\theta$ and similarly for $\k'_\theta$), and that referring to the system in the initial configuration. The scattering amplitude is defined, in the Born approximation\cite{Sakurai}, by
\bq\label{def_f}
f(\k',\k)=-\frac{m_0}{2\pi\hbar^2}\int\D^3\r \,e^{-i(\k'-\k)\cdot\r/\hbar}\,V(\r),
\eq
where $V(\r)$ is the scattering potential and $m_0$ the mass of the environmental particle.
The asymmetric geometry of the system prevents a simple evaluation of the scattering amplitude $f(\k, \k')$ in Eq.~\eqref{def_f}, as well as of the ratio $R(\k,\k')$. Nevertheless we can  provide a reasonable estimate of the decoherence rates. First, we consider as negligible the contributions from the massless rigid bar connecting the two sphere of the nanorod. Thus, the rotational decoherence rate can be simply deduced from the translational one\cite{Schlosshauer:2007aa}
\bqali
 \Lambda_\text{\tiny D}^{(i)}=2\times&\int \D^3\k\int \frac{\D^2\hat {n}'}{4\pi} v(k) \mu(k) |f(\k,k\hat n')|^2\\&\cdot\left(1-e^{\tfrac i\hbar(k\hat n'-\k)\cdot(\r_0^{\theta}-\r_0)}	\right),
\eqali
where the factor 2 takes into account the presence of two spheres, $\r_0^\theta$ corresponding to $\r_0$ rotated by an angle $\theta$ and $R(\k,\k')$ is substituted with the usual phase for translational decoherence.
Next, 
we consider the low temperature regime where we can consider $\braket{\k}\cdot\r_0/\hbar\ll1$, and we obtain
\bqali
 \Lambda_\text{\tiny D}^{(i)}=&\int \D^3\k\int \frac{\D^2\hat {n}'}{4\pi} v(k) \mu(k) |f(\k,k\hat n')|^2\\&\cdot\frac{k^2L^2}{\hbar^2}(\hat n'-\hat n)^2\sin^2\tfrac\theta2,
\eqali
where $\k=k \hat n$ and $(\r_0^{\theta}-\r_0)=2L\sin\tfrac\theta2\times(1,0,0)$.

We have now to determine the differential cross sections for the considered processes. Let us first consider explicitly the case of the collisions with the residual gas. For a spherical system of radius $r$ when the relevant wavelength of the scattering environmental particle is much shorter than the size of the object, the cross section is  equal to the geometrical one\cite{Schlosshauer:2007aa}: $|f(\k',\k)|^2=r^2/4$. 
Therefore the collisional decoherence rate for rotational superpositions becomes: 
\bq\label{coll}
\Lambda_\text{\tiny D}^\text{\tiny coll}=\frac{64n_\text{\tiny gas}\sqrt{2\pi m_\text{\tiny gas}}}{3\hbar^2}r^2L^2(k_\text{\tiny B}T_\text{\tiny E})^{3/2}\sin^2\tfrac\theta2.
\eq
The same applies
 to decoherence  due to environmental thermal photons: scattering, absorption and emission. 
 For the scattering with the thermal photons the cross section is twice that of Ref.\cite{Schlosshauer:2007aa}:
\bq
|f(\k',\k)|^2=\frac{k^4}{\hbar^4}r^6 \Re\left(\frac{\epsilon-1}{\epsilon+2}\right)^2\frac{(1+(\hat n\cdot\hat n')^2)}2,
\eq
where $\epsilon$ is the complex dielectric constant of the system. This expression leads to
\bq\label{scatt-9}
\Lambda_\text{\tiny D}^\text{\tiny scatt}=\frac{64\cdot 8!\xi(9)}{9\pi}r^6L^2c\, \Re\left(\frac{\epsilon-1}{\epsilon+2}\right)^2\left(\frac{k_\text{\tiny B}T_\text{\tiny E}}{\hbar c}\right)^9\sin^2\tfrac\theta2.
\eq
Similarly, for the emission (absorption) of thermal photons the cross section reads
\bq
|f(\k',\k)|^2=4\pi \Im\left(\frac{\epsilon-1}{\epsilon+2}	\right)\frac{k\, r^3}{\hbar},
\eq
from which we obtain
\bq
\Lambda_\text{\tiny D}^\text{\tiny em (ab)}=\frac{128\pi^5}{189}\Im\left(\frac{\epsilon-1}{\epsilon+2}	\right)c\, r^3 L^2 \left(\frac{k_\text{\tiny B}T_\text{\tiny I(E)}}{\hbar c}\right)^6\sin^2\tfrac\theta2,
\eq
where $T_\text{\tiny I}$ is the internal temperature of the system, which in principle can differ from the external temperature $T_\text{\tiny E}$.

For the decoherence due to the scattering with thermal photons an alternative expression was provided in \cite{Zhong:2016aa}:
\bq\label{scatt-7}
\Lambda_\text{\tiny D}^\text{\tiny scatt}=6!\frac{2c}{9\epsilon_0^2}\left(\frac{k_\text{\tiny B}T_\text{\tiny E}}{\hbar c}\right)^7\zeta(7)(\alpha_x-\alpha_z)^2\sin^2\theta,
\eq
where $\alpha_i$ is the $i$-th polarizability component. In the low temperature case here considered, the global decoherence rate computed by using the expression in Eq.~\eqref{scatt-9} does not differ appreciably from that obtained by using the expression in Eq.~\eqref{scatt-7}.


\begin{thebibliography}{29}%
\makeatletter
\providecommand \@ifxundefined [1]{%
 \@ifx{#1\undefined}
}%
\providecommand \@ifnum [1]{%
 \ifnum #1\expandafter \@firstoftwo
 \else \expandafter \@secondoftwo
 \fi
}%
\providecommand \@ifx [1]{%
 \ifx #1\expandafter \@firstoftwo
 \else \expandafter \@secondoftwo
 \fi
}%
\providecommand \natexlab [1]{#1}%
\providecommand \enquote  [1]{``#1''}%
\providecommand \bibnamefont  [1]{#1}%
\providecommand \bibfnamefont [1]{#1}%
\providecommand \citenamefont [1]{#1}%
\providecommand \href@noop [0]{\@secondoftwo}%
\providecommand \href [0]{\begingroup \@sanitize@url \@href}%
\providecommand \@href[1]{\@@startlink{#1}\@@href}%
\providecommand \@@href[1]{\endgroup#1\@@endlink}%
\providecommand \@sanitize@url [0]{\catcode `\\12\catcode `\$12\catcode
  `\&12\catcode `\#12\catcode `\^12\catcode `\_12\catcode `\%12\relax}%
\providecommand \@@startlink[1]{}%
\providecommand \@@endlink[0]{}%
\providecommand \url  [0]{\begingroup\@sanitize@url \@url }%
\providecommand \@url [1]{\endgroup\@href {#1}{\urlprefix }}%
\providecommand \urlprefix  [0]{URL }%
\providecommand \Eprint [0]{\href }%
\providecommand \doibase [0]{http://dx.doi.org/}%
\providecommand \selectlanguage [0]{\@gobble}%
\providecommand \bibinfo  [0]{\@secondoftwo}%
\providecommand \bibfield  [0]{\@secondoftwo}%
\providecommand \translation [1]{[#1]}%
\providecommand \BibitemOpen [0]{}%
\providecommand \bibitemStop [0]{}%
\providecommand \bibitemNoStop [0]{.\EOS\space}%
\providecommand \EOS [0]{\spacefactor3000\relax}%
\providecommand \BibitemShut  [1]{\csname bibitem#1\endcsname}%
\let\auto@bib@innerbib\@empty
%</preamble>
\bibitem [{\citenamefont {Feynman}(2002)}]{FeynmanGravitation}%
  \BibitemOpen
  \bibfield  {author} {\bibinfo {author} {\bibfnamefont {R.~P.}\ \bibnamefont
  {Feynman}},\ }\href@noop {} {\emph {\bibinfo {title} {Feynman Lectures on
  Gravitation}}},\ edited by\ \bibinfo {editor} {\bibfnamefont {R.~P.}\
  \bibnamefont {Feynman}}, \bibinfo {editor} {\bibfnamefont {F.~B.}\
  \bibnamefont {Morinigo}}, \bibinfo {editor} {\bibfnamefont {W.}~\bibnamefont
  {Wagner}}, \ and\ \bibinfo {editor} {\bibfnamefont {B.}~\bibnamefont
  {Hatfield}}\ (\bibinfo  {publisher} {Avalon Publishing},\ \bibinfo {year}
  {2002})\BibitemShut {NoStop}%
\bibitem [{\citenamefont {Penrose}(1996)}]{PenroseCollapse}%
  \BibitemOpen
  \bibfield  {author} {\bibinfo {author} {\bibfnamefont {R.}~\bibnamefont
  {Penrose}},\ }\href {\doibase 10.1007/BF02105068} {\bibfield  {journal}
  {\bibinfo  {journal} {General Relativity and Gravitation}\ }\textbf {\bibinfo
  {volume} {28}},\ \bibinfo {pages} {581} (\bibinfo {year} {1996})}\BibitemShut
  {NoStop}%
\bibitem [{\citenamefont {Hossenfelder}(2011)}]{hoss}%
  \BibitemOpen
  \bibfield  {author} {\bibinfo {author} {\bibfnamefont {S.}~\bibnamefont
  {Hossenfelder}},\ }\href@noop {} {\emph {\bibinfo {title} {Classical and
  Quantum Gravity: Theory, Analysis and Applications}}},\ edited by\ \bibinfo
  {editor} {\bibfnamefont {V.~R.}\ \bibnamefont {Frignanni}}\ (\bibinfo
  {publisher} {Nova Publishers},\ \bibinfo {year} {2011})\BibitemShut {NoStop}%
\bibitem [{\citenamefont {Ashoorioon}\ \emph {et~al.}(2014)\citenamefont
  {Ashoorioon}, \citenamefont {Dev},\ and\ \citenamefont {Mazumdar}}]{arun}%
  \BibitemOpen
  \bibfield  {author} {\bibinfo {author} {\bibfnamefont {A.}~\bibnamefont
  {Ashoorioon}}, \bibinfo {author} {\bibfnamefont {P.~S.~B.}\ \bibnamefont
  {Dev}}, \ and\ \bibinfo {author} {\bibfnamefont {A.}~\bibnamefont
  {Mazumdar}},\ }\href
  {http://www.worldscientific.com/doi/abs/10.1142/S0217732314501636} {\bibfield
   {journal} {\bibinfo  {journal} {Mod.~Phys.~Lett.~A}\ }\textbf {\bibinfo
  {volume} {29}},\ \bibinfo {pages} {1450163} (\bibinfo {year}
  {2014})}\BibitemShut {NoStop}%
\bibitem [{\citenamefont {Kiefer}(2006)}]{kiefer}%
  \BibitemOpen
  \bibfield  {author} {\bibinfo {author} {\bibfnamefont {C.}~\bibnamefont
  {Kiefer}},\ }\href {http://dx.doi.org/10.1002/andp.200510175} {\bibfield
  {journal} {\bibinfo  {journal} {Ann.~Phys.}\ }\textbf {\bibinfo {volume}
  {15}},\ \bibinfo {pages} {129} (\bibinfo {year} {2006})}\BibitemShut
  {NoStop}%
\bibitem [{\citenamefont {Bassi}\ \emph {et~al.}(2017)\citenamefont {Bassi},
  \citenamefont {Gro{\ss}ardt},\ and\ \citenamefont {Ulbricht}}]{bassi}%
  \BibitemOpen
  \bibfield  {author} {\bibinfo {author} {\bibfnamefont {A.}~\bibnamefont
  {Bassi}}, \bibinfo {author} {\bibfnamefont {A.}~\bibnamefont {Gro{\ss}ardt}},
  \ and\ \bibinfo {author} {\bibfnamefont {H.}~\bibnamefont {Ulbricht}},\
  }\href {http://stacks.iop.org/0264-9381/34/i=19/a=193002} {\bibfield
  {journal} {\bibinfo  {journal} {Class. Quantum Grav.}\ }\textbf {\bibinfo
  {volume} {34}},\ \bibinfo {pages} {193002} (\bibinfo {year}
  {2017})}\BibitemShut {NoStop}%
\bibitem [{\citenamefont {Bahrami}\ \emph {et~al.}(2015)\citenamefont
  {Bahrami}, \citenamefont {Bassi}, \citenamefont {McMillen}, \citenamefont
  {Paternostro},\ and\ \citenamefont {Ulbricht}}]{bahrami}%
  \BibitemOpen
  \bibfield  {author} {\bibinfo {author} {\bibfnamefont {M.}~\bibnamefont
  {Bahrami}}, \bibinfo {author} {\bibfnamefont {A.}~\bibnamefont {Bassi}},
  \bibinfo {author} {\bibfnamefont {S.}~\bibnamefont {McMillen}}, \bibinfo
  {author} {\bibfnamefont {M.}~\bibnamefont {Paternostro}}, \ and\ \bibinfo
  {author} {\bibfnamefont {H.}~\bibnamefont {Ulbricht}},\ }\href
  {https://arxiv.org/abs/1507.05733} {\bibfield  {journal} {\bibinfo  {journal}
  {ArXiv}\ } (\bibinfo {year} {2015})},\ \Eprint
  {http://arxiv.org/abs/1507.05733} {1507.05733} \BibitemShut {NoStop}%
\bibitem [{\citenamefont {Kafri}\ \emph {et~al.}(2014)\citenamefont {Kafri},
  \citenamefont {Taylor},\ and\ \citenamefont {Milburn}}]{kafri}%
  \BibitemOpen
  \bibfield  {author} {\bibinfo {author} {\bibfnamefont {D.}~\bibnamefont
  {Kafri}}, \bibinfo {author} {\bibfnamefont {J.~M.}\ \bibnamefont {Taylor}}, \
  and\ \bibinfo {author} {\bibfnamefont {G.~J.}\ \bibnamefont {Milburn}},\
  }\href {http://stacks.iop.org/1367-2630/16/i=6/a=065020} {\bibfield
  {journal} {\bibinfo  {journal} {New J.~Phys.}\ }\textbf {\bibinfo {volume}
  {16}},\ \bibinfo {pages} {065020} (\bibinfo {year} {2014})}\BibitemShut
  {NoStop}%
\bibitem [{\citenamefont {Boys}(1894)}]{Cavendish}%
  \BibitemOpen
  \bibfield  {author} {\bibinfo {author} {\bibfnamefont {C.~V.}\ \bibnamefont
  {Boys}},\ }\href@noop {} {\bibfield  {journal} {\bibinfo  {journal} {Nature}\
  }\textbf {\bibinfo {volume} {50}} (\bibinfo {year} {1894})}\BibitemShut
  {NoStop}%
\bibitem [{\citenamefont {Scala}\ \emph {et~al.}(2013)\citenamefont {Scala}
  \emph {et~al.}}]{Scala:2013aa}%
  \BibitemOpen
  \bibfield  {author} {\bibinfo {author} {\bibfnamefont {M.}~\bibnamefont
  {Scala}} \emph {et~al.},\ }\href
  {http://link.aps.org/doi/10.1103/PhysRevLett.111.180403} {\bibfield
  {journal} {\bibinfo  {journal} {Phys.~Rev.~Lett.}\ }\textbf {\bibinfo
  {volume} {111}},\ \bibinfo {pages} {180403} (\bibinfo {year}
  {2013})}\BibitemShut {NoStop}%
\bibitem [{\citenamefont {Arita}\ \emph {et~al.}(2013)\citenamefont {Arita},
  \citenamefont {Mazilu},\ and\ \citenamefont {Dholakia}}]{Arita2013}%
  \BibitemOpen
  \bibfield  {author} {\bibinfo {author} {\bibfnamefont {Y.}~\bibnamefont
  {Arita}}, \bibinfo {author} {\bibfnamefont {M.}~\bibnamefont {Mazilu}}, \
  and\ \bibinfo {author} {\bibfnamefont {K.}~\bibnamefont {Dholakia}},\ }\href
  {http://dx.doi.org/10.1038/ncomms3374} {\bibfield  {journal} {\bibinfo
  {journal} {Nat.~Commun.}\ }\textbf {\bibinfo {volume} {4}} (\bibinfo {year}
  {2013})}\BibitemShut {NoStop}%
\bibitem [{\citenamefont {Kuhn}\ \emph {et~al.}(2015)\citenamefont {Kuhn} \emph
  {et~al.}}]{Kuhn2015}%
  \BibitemOpen
  \bibfield  {author} {\bibinfo {author} {\bibfnamefont {S.}~\bibnamefont
  {Kuhn}} \emph {et~al.},\ }\href
  {http://dx.doi.org/10.1021/acs.nanolett.5b02302} {\bibfield  {journal}
  {\bibinfo  {journal} {Nano Lett.}\ }\textbf {\bibinfo {volume} {15}},\
  \bibinfo {pages} {5604} (\bibinfo {year} {2015})}\BibitemShut {NoStop}%
\bibitem [{\citenamefont {Kuhn}\ \emph
  {et~al.}(2017{\natexlab{a}})\citenamefont {Kuhn} \emph {et~al.}}]{Kuhn2017}%
  \BibitemOpen
  \bibfield  {author} {\bibinfo {author} {\bibfnamefont {S.}~\bibnamefont
  {Kuhn}} \emph {et~al.},\ }\href
  {http://www.osapublishing.org/optica/abstract.cfm?URI=optica-4-3-356}
  {\bibfield  {journal} {\bibinfo  {journal} {Optica}\ }\textbf {\bibinfo
  {volume} {4}},\ \bibinfo {pages} {356} (\bibinfo {year}
  {2017}{\natexlab{a}})}\BibitemShut {NoStop}%
\bibitem [{\citenamefont {Hoang}\ \emph {et~al.}(2016)\citenamefont {Hoang}
  \emph {et~al.}}]{Hoang2016}%
  \BibitemOpen
  \bibfield  {author} {\bibinfo {author} {\bibfnamefont {T.~M.}\ \bibnamefont
  {Hoang}} \emph {et~al.},\ }\href
  {https://link.aps.org/doi/10.1103/PhysRevLett.117.123604} {\bibfield
  {journal} {\bibinfo  {journal} {Phys.~Rev.~Lett.}\ }\textbf {\bibinfo
  {volume} {117}},\ \bibinfo {pages} {123604} (\bibinfo {year}
  {2016})}\BibitemShut {NoStop}%
\bibitem [{\citenamefont {Rahman}\ and\ \citenamefont
  {Barker}(2017)}]{Rahman2017}%
  \BibitemOpen
  \bibfield  {author} {\bibinfo {author} {\bibfnamefont {A.~T.~M.}\
  \bibnamefont {Rahman}}\ and\ \bibinfo {author} {\bibfnamefont {P.~F.}\
  \bibnamefont {Barker}},\ }\href {https://doi.org/10.1038/s41566-017-0005-3}
  {\bibfield  {journal} {\bibinfo  {journal} {Nature Phot.}\ }\textbf {\bibinfo
  {volume} {11}} (\bibinfo {year} {2017})}\BibitemShut {NoStop}%
\bibitem [{\citenamefont {Jain}\ \emph {et~al.}(2016)\citenamefont {Jain} \emph
  {et~al.}}]{jain2016direct}%
  \BibitemOpen
  \bibfield  {author} {\bibinfo {author} {\bibfnamefont {V.}~\bibnamefont
  {Jain}} \emph {et~al.},\ }\href
  {https://link.aps.org/doi/10.1103/PhysRevLett.116.243601} {\bibfield
  {journal} {\bibinfo  {journal} {Phys.~Rev.~Lett.}\ }\textbf {\bibinfo
  {volume} {116}},\ \bibinfo {pages} {243601} (\bibinfo {year}
  {2016})}\BibitemShut {NoStop}%
\bibitem [{\citenamefont {Kuhn}\ \emph
  {et~al.}(2017{\natexlab{b}})\citenamefont {Kuhn} \emph {et~al.}}]{Kuhn2017b}%
  \BibitemOpen
  \bibfield  {author} {\bibinfo {author} {\bibfnamefont {S.}~\bibnamefont
  {Kuhn}} \emph {et~al.},\ }\href {https://arxiv.org/abs/1702.07565} {\bibfield
   {journal} {\bibinfo  {journal} {ArXiv}\ } (\bibinfo {year}
  {2017}{\natexlab{b}})},\ \Eprint {http://arxiv.org/abs/1702.07565}
  {1702.07565} \BibitemShut {NoStop}%
\bibitem [{\citenamefont {Levitt}(2001)}]{Levitt2001}%
  \BibitemOpen
  \bibfield  {author} {\bibinfo {author} {\bibfnamefont {M.}~\bibnamefont
  {Levitt}},\ }\href@noop {} {\emph {\bibinfo {title} {Spin Dynamics: Basics of
  Nuclear Magnetic Resonance}}}\ (\bibinfo  {publisher} {John Wiley \& Sons},\
  \bibinfo {year} {2001})\BibitemShut {NoStop}%
\bibitem [{\citenamefont {Andrich}\ \emph {et~al.}(2014)\citenamefont {Andrich}
  \emph {et~al.}}]{Andrich}%
  \BibitemOpen
  \bibfield  {author} {\bibinfo {author} {\bibfnamefont {P.}~\bibnamefont
  {Andrich}} \emph {et~al.},\ }\href {http://dx.doi.org/10.1021/nl501208s}
  {\bibfield  {journal} {\bibinfo  {journal} {Nano Lett.}\ }\textbf {\bibinfo
  {volume} {14}},\ \bibinfo {pages} {4959} (\bibinfo {year}
  {2014})}\BibitemShut {NoStop}%
\bibitem [{\citenamefont {Trusheim}\ \emph {et~al.}(2014)\citenamefont
  {Trusheim} \emph {et~al.}}]{Trusheim2014}%
  \BibitemOpen
  \bibfield  {author} {\bibinfo {author} {\bibfnamefont {M.~E.}\ \bibnamefont
  {Trusheim}} \emph {et~al.},\ }\href {http://dx.doi.org/10.1021/nl402799u}
  {\bibfield  {journal} {\bibinfo  {journal} {Nano Lett.}\ }\textbf {\bibinfo
  {volume} {14}},\ \bibinfo {pages} {32} (\bibinfo {year} {2014})}\BibitemShut
  {NoStop}%
\bibitem [{\citenamefont {Schirhagl}\ \emph {et~al.}(2014)\citenamefont
  {Schirhagl}, \citenamefont {Chang}, \citenamefont {Loretz},\ and\
  \citenamefont {Degen}}]{Schirhagl}%
  \BibitemOpen
  \bibfield  {author} {\bibinfo {author} {\bibfnamefont {R.}~\bibnamefont
  {Schirhagl}}, \bibinfo {author} {\bibfnamefont {K.}~\bibnamefont {Chang}},
  \bibinfo {author} {\bibfnamefont {M.}~\bibnamefont {Loretz}}, \ and\ \bibinfo
  {author} {\bibfnamefont {C.~L.}\ \bibnamefont {Degen}},\ }\href
  {https://doi.org/10.1146/annurev-physchem-040513-103659} {\bibfield
  {journal} {\bibinfo  {journal} {Annu.~Rev.~Phys.~Chem.}\ }\textbf {\bibinfo
  {volume} {65}},\ \bibinfo {pages} {83} (\bibinfo {year} {2014})}\BibitemShut
  {NoStop}%
\bibitem [{\citenamefont {Mamin}\ \emph {et~al.}(2007)\citenamefont {Mamin}
  \emph {et~al.}}]{Mamin}%
  \BibitemOpen
  \bibfield  {author} {\bibinfo {author} {\bibfnamefont {H.}~\bibnamefont
  {Mamin}} \emph {et~al.},\ }\href@noop {} {\bibfield  {journal} {\bibinfo
  {journal} {Nature Nanotech.}\ }\textbf {\bibinfo {volume} {2}} (\bibinfo
  {year} {2007})}\BibitemShut {NoStop}%
\bibitem [{\citenamefont {Vovrosh}\ \emph {et~al.}(2017)\citenamefont {Vovrosh}
  \emph {et~al.}}]{Vovrosh2017}%
  \BibitemOpen
  \bibfield  {author} {\bibinfo {author} {\bibfnamefont {J.}~\bibnamefont
  {Vovrosh}} \emph {et~al.},\ }\href
  {http://josab.osa.org/abstract.cfm?URI=josab-34-7-1421} {\bibfield  {journal}
  {\bibinfo  {journal} {J.~Opt.~Soc.~Am.~B}\ }\textbf {\bibinfo {volume}
  {34}},\ \bibinfo {pages} {1421} (\bibinfo {year} {2017})}\BibitemShut
  {NoStop}%
\bibitem [{\citenamefont {Aspelmeyer}\ \emph {et~al.}(2014)\citenamefont
  {Aspelmeyer}, \citenamefont {Kippenberg},\ and\ \citenamefont
  {Marquardt}}]{Aspelmeyer2014}%
  \BibitemOpen
  \bibfield  {author} {\bibinfo {author} {\bibfnamefont {M.}~\bibnamefont
  {Aspelmeyer}}, \bibinfo {author} {\bibfnamefont {T.~J.}\ \bibnamefont
  {Kippenberg}}, \ and\ \bibinfo {author} {\bibfnamefont {F.}~\bibnamefont
  {Marquardt}},\ }\href {https://link.aps.org/doi/10.1103/RevModPhys.86.1391}
  {\bibfield  {journal} {\bibinfo  {journal} {Rev. Mod. Phys.}\ }\textbf
  {\bibinfo {volume} {86}},\ \bibinfo {pages} {1391} (\bibinfo {year}
  {2014})}\BibitemShut {NoStop}%
\bibitem [{\citenamefont {Wan}\ \emph {et~al.}(2016)\citenamefont {Wan} \emph
  {et~al.}}]{Wan2016}%
  \BibitemOpen
  \bibfield  {author} {\bibinfo {author} {\bibfnamefont {C.}~\bibnamefont
  {Wan}} \emph {et~al.},\ }\href
  {https://link.aps.org/doi/10.1103/PhysRevLett.117.143003} {\bibfield
  {journal} {\bibinfo  {journal} {Phys.~Rev.~Lett.}\ }\textbf {\bibinfo
  {volume} {117}},\ \bibinfo {pages} {143003} (\bibinfo {year}
  {2016})}\BibitemShut {NoStop}%
\bibitem [{\citenamefont {Schlosshauer}(2007)}]{Schlosshauer:2007aa}%
  \BibitemOpen
  \bibfield  {author} {\bibinfo {author} {\bibfnamefont {M.~A.}\ \bibnamefont
  {Schlosshauer}},\ }\href@noop {} {\emph {\bibinfo {title} {{Decoherence and
  the Quantum-To-Classical Transition}}}},\ \bibinfo {edition} {1st}\ ed.\
  (\bibinfo  {publisher} {Springer-Verlag Berlin Heidelberg},\ \bibinfo {year}
  {2007})\BibitemShut {NoStop}%
\bibitem [{\citenamefont {Romero-Isart}(2011)}]{Romero-Isart:2011aa}%
  \BibitemOpen
  \bibfield  {author} {\bibinfo {author} {\bibfnamefont {O.}~\bibnamefont
  {Romero-Isart}},\ }\href {\doibase 10.1103/PhysRevA.84.052121} {\bibfield
  {journal} {\bibinfo  {journal} {Phys.~Rev.~A}\ }\textbf {\bibinfo {volume}
  {84}},\ \bibinfo {pages} {052121} (\bibinfo {year} {2011})}\BibitemShut
  {NoStop}%
\bibitem [{\citenamefont {Zhong}\ and\ \citenamefont
  {Robicheaux}(2016)}]{Zhong:2016aa}%
  \BibitemOpen
  \bibfield  {author} {\bibinfo {author} {\bibfnamefont {C.}~\bibnamefont
  {Zhong}}\ and\ \bibinfo {author} {\bibfnamefont {F.}~\bibnamefont
  {Robicheaux}},\ }\href {http://link.aps.org/doi/10.1103/PhysRevA.94.052109}
  {\bibfield  {journal} {\bibinfo  {journal} {Phys.~Rev.~A}\ }\textbf {\bibinfo
  {volume} {94}},\ \bibinfo {pages} {052109} (\bibinfo {year}
  {2016})}\BibitemShut {NoStop}%
\bibitem [{\citenamefont {Sakurai}\ and\ \citenamefont
  {Napolitano}(2011)}]{Sakurai}%
  \BibitemOpen
  \bibfield  {author} {\bibinfo {author} {\bibfnamefont {J.~J.}\ \bibnamefont
  {Sakurai}}\ and\ \bibinfo {author} {\bibfnamefont {J.}~\bibnamefont
  {Napolitano}},\ }\href@noop {} {\emph {\bibinfo {title} {Modern Quantum
  Mechanics}}}\ (\bibinfo  {publisher} {Addison-Wesley},\ \bibinfo {year}
  {2011})\BibitemShut {NoStop}%
\end{thebibliography}
\end{document}